\documentclass{iau} 
\usepackage{graphicx}
\usepackage{bm,url}

\newcommand{\mps}{m~s$^{-1}$}

\newcommand{\cmss}{cm$^2$~s$^{-1}$}

\newcommand{\Fig}[1]{Figure~\ref{#1}}

\newcommand{\brac}[1]{\langle #1 \rangle}

\newcommand\mnras{{MNRAS}}%
\newcommand\aap{{Astrophys. J}}%
\newcommand\apjl{{Astrophys. J}}%
\newcommand\apj{{Astrophys. J}}%
\newcommand\solphys{{Sol Phys}}%
\newcommand\ssr{{Space~Sci.~Rev.}}%

\title[Solar cycle variability in Babcock-Leighton dynamo] 
{Long-term variability of the solar cycle in the Babcock-Leighton dynamo framework}

\author[B.\ B. Karak \& M. Miesch]   
{Bidya Binay Karak$^1$
 \and Mark Miesch$^2$}

\affiliation{$^1$Indian Institute of Astrophysics, Koramangala, Bangalore 560034, India \\ email: {\tt bidyakarak@gmail.com} \\[\affilskip]
$^2$National Center for Atmospheric Research, 3080 Center Green Dr., Boulder, CO 80301, USA \\email: {\tt miesch@ucar.edu}}

\pubyear{2015}
\volume{xxx}  
\jname{Title of your IAU Symposium}
\editors{A.C. Editor, B.D. Editor \& C.E. Editor, eds.}
\begin{document}

\maketitle

\begin{abstract}

We explore the cause of the solar cycle variabilities using a novel 3D Babcock--Leighton dynamo model.  
In this model, based on the toroidal flux at the base of the convection zone, bipolar magnetic regions (BMRs) are produced with statistical properties
obtained from observed distributions. 
We find that a little quenching in BMR tilt is sufficient to stabilize the dynamo growth. The randomness and nonlinearity in the BMR emergences make the poloidal field unequal and cause some variability in the solar cycle. 
However, when observed scatter of BMR tilts around Joy's law with a standard deviation of $15^\circ$, is considered, 
our model produces a variation in the solar cycle, including north-south
asymmetry comparable to the observations.
The morphology of magnetic fields 
closely resembles observations, in particular
the surface radial field possesses a more mixed
polarity field.
Observed scatter also produces grand minima. In 11,650 years of simulation,
17 grand minima are detected and $11\%$ of its time the model remained in these grand minima. 
When we double the tilt scatter, the model produces correct statistics of grand minima. Importantly, 
the dynamo continues even during grand minima with only a few BMRs, without requiring any additional alpha effect. 
The reason for this is the downward magnetic pumping which suppresses 
the diffusion of the magnetic flux across the surface.
The magnetic pumping also helps to achieve 11-year magnetic cycle
using the observed BMR flux distribution, even at the high diffusivity. 

\end{abstract}

From last 400 years of sunspot observations, 
we know that the solar cycle is irregular. 
The individual cycle strength and duration vary cycle-to-cycle. 
The extreme example of this irregularity is the Maunder minimum in the 17th century when sunspot cycle went to a very low value.  
We now know that Sun had many such events in the past (\cite{USK07}).
Here we discuss the cause of this irregular solar cycle using 
a Babcock-Leighton (BL) dynamo model in which the poloidal field is 
generated near the surface through the decay and dispersal of 
tilted bipolar active regions (BMRs)---the BL process (\cite{Kar14a}). 
We note that recent magnetohydrodynamics 
simulations have also started to produce irregular magnetic cycles, 
including grand minima (\cite{ABMT15,KKB15,Kap16}), but here we focus only 
on the BL dynamo.
 
The generation of poloidal field in BL process relies on the tilt of BMR. 
From observations, we know that while the mean tilt of BMRs is governed by 
Joy's law, there is a considerable scatter around it (\cite{Das10,SK12}). 
This scatter can alter the poloidal field at each solar minimum from its average value
(e.g., \cite{JCS14,HCM17}). 
Based on this idea, 
previous authors  have already included fluctuations in the poloidal field source term in their axisymmetric dynamo models and have produced variable solar cycle including Maunder-like grand minima 
(Choudhuri et al.\ 2007; Choudhuri \& Karak 2009, 2012; 
Karak 2010; Karak \& Choudhuri 2011, 2012, 2013; Passos et al.\ 2014). 
However, due to the limitation of previous axisymmetric models, explicit BMRs with actual tilts 
were not included in those models. Recently, \cite{LC16} coupled 
a 2D surface flux transport model with a 2D 
flux transport dynamo model, operating in the CZ and they 
have shown that the observed tilt scatter produces variable 
solar cycle including the Dalton-like grand minima.

On the other hand, \cite{MD14} and \cite{MT16} have recently developed a novel 3D dynamo model,
in which explicit tilted BMRs are deposited based on the toroidal flux at the base of the CZ. 
Later, Karak \& Miesch (2017), hereafter KM17, have improved this model by incorporating statistical features of BMRs from observations. 
Here we present a summary of our previous studies (KM17, \cite{KM18}) in which 
we have demonstrated that 
the scatter in the sunspot tilt can produce variability in the solar cycle,
including grand minima.
The mechanism of the recovery of grand minima under the BL process alone
is also highlighted.

For the dyamo model used in our study, 
we refer the readers to our previous publication KM17.
From this publication, we consider Runs~B9--11 in which the diffusivity near the surface is $4.5\times10^{12}$~\cmss and in the bulk of CZ, it is $1.5\times10^{12}$~\cmss.  
The BMR flux distribution is fixed at the observed value
and the rate of BMR eruption changes in response to the toroidal flux at the BCZ.
The BMR tilt has a Gaussian scatter around Joy's law
with standard deviation $\sigma_\delta$.
A downward magnetic pumping with a speed of $20$~\mps\ near the surface
is also included in this model.

\begin{figure}
\begin{minipage}[t]{0.38\textwidth}
\centerline{\includegraphics[width=.9\columnwidth]{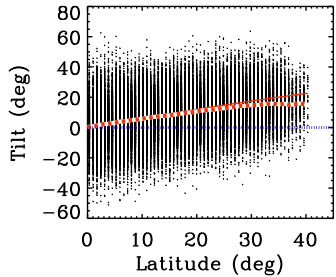}}
\end{minipage}%
\begin{minipage}[t]{0.60\textwidth}
\vspace{-3.6cm}
\caption{
Results obtained from dynamo simulation with $\sigma_\delta = 15^\circ$ (Run~B10 of KM17): 
BMR tilts versus latitudes ($\lambda$) shown only from the northern hemisphere data. The
solid, dashed, and dotted lines respectively show the actual Joy's law: 
$\delta = 35^\circ \sin \lambda$, the mean BMR tilts in each latitudes, 
and the zero line. 
Note that the dashed line deviates from the Joy's law (solid line) 
due to the nonlinear quenching introduced in it; see Equation 10 of KM17 for details.
}
\end{minipage}%
\end{figure}

\begin{figure}
\centerline{\includegraphics[width=1.0\columnwidth]{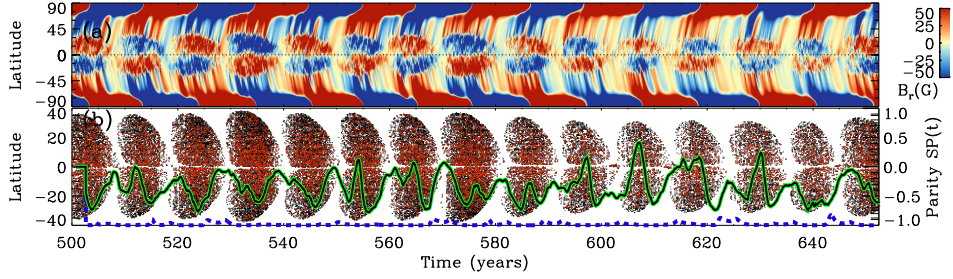}}
\caption{
Results from simulation with $\sigma_\delta = 15^\circ$ (Run~B10 of KM17): 
temporal variations of (a) azimuthal-averaged surface radial field $\brac{B_r(R,\theta,\phi)}_\phi$,
and (b) latitudes of BMRs; red points show the wrongly tilted BMRs; 
green/solid and blue/dashed lines show symmetric parities, 
computed over the four years of
surface $B_r$ and the bottom $B_\phi$, respectively.
}
\label{fig:bfly}
\end{figure}
\begin{figure*}
\centering
\includegraphics[width=1.0\columnwidth]{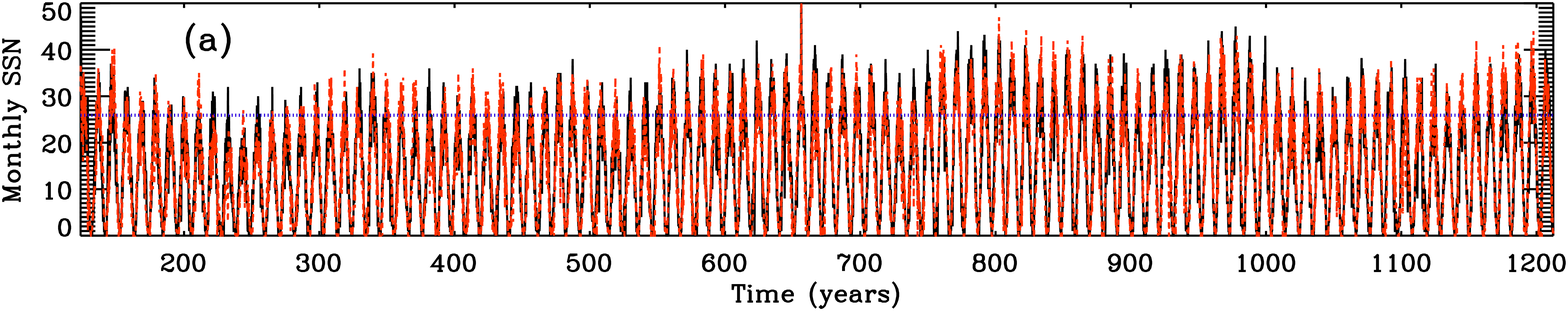}
\includegraphics[width=1.0\columnwidth]{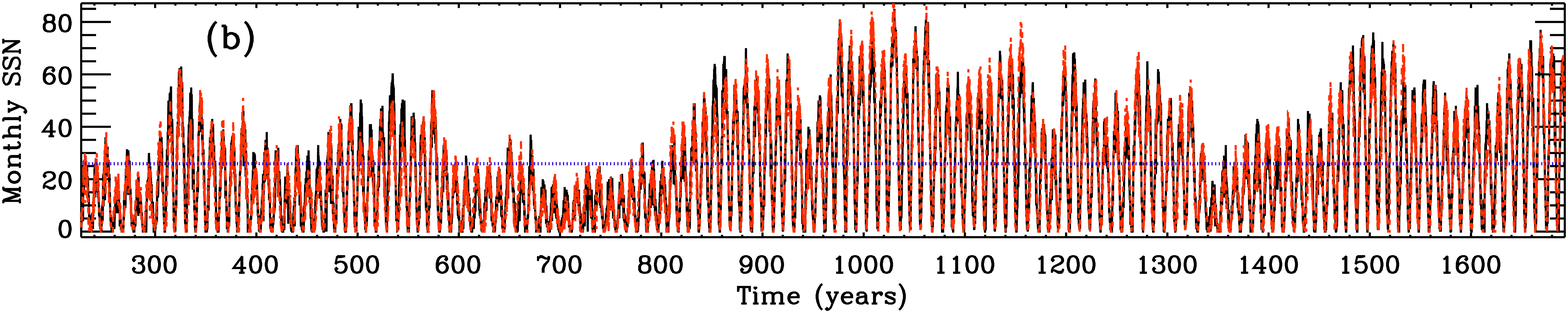}
\includegraphics[width=1.0\columnwidth]{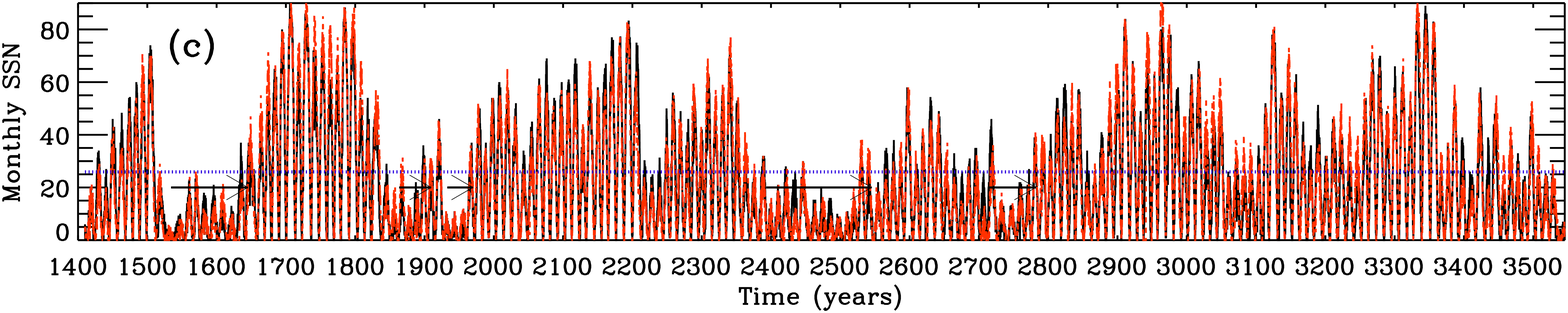}
\caption{Time series of the monthly sunspot number (SSN) (which is same as the BMR number) 
from the simulation (a) without tilt scatter (Run~B9), with scatter of (b) $\sigma_\delta=15^\circ$ (Run B10), and (c) $\sigma_\delta=30^\circ$ (Run B11).
The horizontal line shows the mean of peaks of the monthly BMRs obtained for last $13$ 
observed solar cycles.
Arrows in (c) represent the locations of grand minima.
}
\label{fig:ssn}
\end{figure*}

A few cycles from the dynamo simulation with tilt scatter of $\sigma_\delta = 15^\circ$ 
around Joy's law is shown in \Fig{fig:bfly}, while the BMR number 
from a longer simulation is shown in \Fig{fig:ssn} (b). 
Although there is a considerable variation
in the magnetic field, the model produces an overall stable magnetic cycle 
due to a nonlinear quenching in the tilt angle. Interestingly, as seen in Figure~1, 
to stabilize the dynamo growth,
only a little quenching is needed.

In \Fig{fig:bfly} we find that most of the basic features of 
solar magnetic cycle, 
namely, 11 years 
periodicity, regular polarity reversal, and equatorward migration of sunspots,
and poleward migration of radial field on the surface 
are reproduced in this model.
We also observe frequently mixed polarity surface field (\Fig{fig:bfly}(a))
which is a consequence of the wrong tilt (Figure~1). 
A significant hemispheric asymmetry in the magnetic field and sunspot number is observed. All these features are broadly consistent with observations (e.g., \cite{McInt13,MKB17}).

We note that despite using much higher diffusivity,
our model produces a dipolar solution. However, the 
magnetic field is largely antisymmetric across the equator (dipolar) 
only near the solar minimum (\Fig{fig:bfly}).
The mean parity of
surface radial field deviates
most strongly from the antisymmetric mode during
cycle maxima. This is again consistent with the solar data (\cite{DBH12}).

We also note that the sunspot number obtained in our model is the
actual spot number produced by the model (\Fig{fig:ssn})
and it is not a proxy as in the previous models (e.g., \cite{Kar14a}). 
The number of BMRs obtained from this model is  consistent with 
the observed value obtained from last 13 observed solar cycles.
Variability of the BMR cycle maxima obtained from this simulation is
$35\%$, in comparison, the variability seen in the observed cycles 
between 1755 and 2008 is $32\%$.
We note that the simulation without tilt scatter also produces a variability
of $11\%$ due to randomness in BMR flux and time delay distributions and 
the nonlinearities in the BL process (\Fig{fig:ssn}(a)).

\begin{figure}
\centerline{ \includegraphics[width=1.\columnwidth]{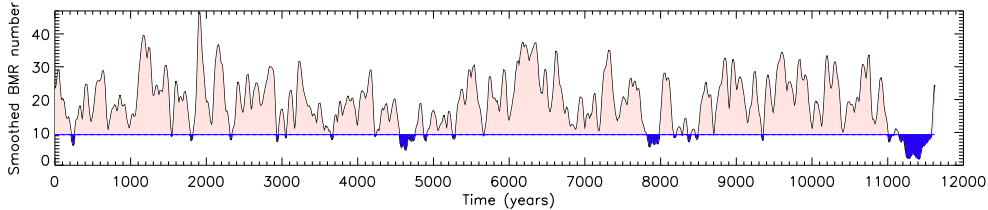}}
\caption{The smoothed BMR number from 11650-year simulation 
with $\sigma_\delta=15^\circ$ (Run B10). Blue shaded regions below $50\%$ of the mean represent the grand minima.}
\label{fig:gm}
\end{figure}

Our model with $\sigma_\delta=15^\circ$ tilt scatter also produces
grand minima as seen in \Fig{fig:gm}. In 11,650 years of simulations,
we have found 17 grand minima. In comparison, 27 grand minima were detected in the solar proxy data 
(\cite{USK07}).
Our model remains in grand minima for $11\%$ of its time, while
this number for the Sun is $17\%$.

When we double the tilt fluctuations, i.e., 
$\sigma_\delta = 30^\circ$, the model produces a correct number of grand minima; see Karak \& Miesch (2018) for details.
Interestingly, the model manages to produce stable cycle
event at this large tilt fluctuations. Also every time, 
the model successfully recovers to the normal cycle.
This was a very unexpected result, because during grand minima when
sunspot number goes to a low value for a few years, 
the generation of poloidal field becomes negligible and we expected
that the model to decay through the turbulent diffusion.
We realized that our model does not allow to shut down 
because of downward magnetic pumping that is included near the surface 
(\cite{KM18}).
The pumping makes the magnetic field radial near the surface and suppresses the diffuse of the horizontal magnetic field across the surface. 
Hence during grand minima when there are no sufficient number of BMRs,
the weak poloidal flux is continuously
be able to produce toroidal flux and eventually, allows
the model to recover from grand minima.



\begin{thebibliography}{}
\bibitem[{{Karak} {et~al.} 2014}]{Kar14a}
{Karak}, B.\ B., {Jiang}, J., {Miesch}, M.\ S., {Charbonneau}, P. et al.\ 2014, \ssr, 186, 561

%
\bibitem[{{Karak} 2010}]{Kar10}
{Karak}, B.~B. 2010, \apj, 724, 1021

\bibitem[{{Karak} \& {Choudhuri} 2011}]{KC11}
{Karak}, B.~B., \& {Choudhuri}, A.~R. 2011, \mnras, 410, 1503

\bibitem[{{Karak} \& {Choudhuri} 2012}]{KC12}
---. 2012, \solphys, 278, 137

\bibitem[{{Karak} \& {Choudhuri} 2013}]{KC13}
---. 2013, Res. Astron. Astrophys., 13, 1339

\bibitem[{{Karak} \& {Cameron} 2016}]{KC16}
{Karak}, B.~B., \& {Cameron}, R. 2016, \apj, 832, 94

\bibitem[{{DeRosa} {et~al.} 2012}]{DBH12}
{DeRosa}, M.~L., {Brun}, A.~S., \& {Hoeksema}, J.~T. 2012, \apj, 757, 96

%
\bibitem[{{Usoskin} {et~al.} 2007}]{USK07}
{Usoskin}, I.~G., {Solanki}, S.~K., \& {Kovaltsov}, G.~A. 2007, \aap, 471, 301

\bibitem[{{Choudhuri} {et~al.} 2007}]{CCJ07}
{Choudhuri}, A.~R., {Chatterjee}, P., \& {Jiang}, J. 2007, Physical Review
  Letters, 98, 131103

\bibitem[{{Jiang} {et~al.} 2014}]{JCS14}
{Jiang}, J., {Cameron}, R.~H., \& {Sch{\"u}ssler}, M. 2014, \apj, 791, 5

\bibitem[{{Choudhuri} \& {Karak} 2009}]{CK09}
{Choudhuri}, A.~R., \& {Karak}, B.~B. 2009, Res. Astron. Astrophys., 9, 953

\bibitem[{{Choudhuri} \& {Karak} 2012}]{CK12}
{Choudhuri}, A.~R., \& {Karak}, B.~B. 2012, Phys. Rev. Lett., 109, 171103

\bibitem[{{Dasi-Espuig} {et~al.} 2010}]{Das10}
{Dasi-Espuig}, M., {Solanki}, S.~K., {Krivova}, N.~A., {Cameron}, R. et al.\ 2010, \aap, 518, A7

\bibitem[{{Miesch} \& {Dikpati}(2014)}]{MD14}
{Miesch}, M.~S., \& {Dikpati}, M. 2014, \apjl, 785, L8

\bibitem[{{Miesch} \& {Teweldebirhan} (2016)}]{MT16}
{Miesch}, M.~S., \& {Teweldebirhan}, K. 2016, \ssr

\bibitem[{{Karak} \& {Miesch} 2017}]{KM17}
{Karak}, B.~B., \& {Miesch}, M.~S., 2017, \apj, 847, 69

\bibitem[{{Karak} \& {Miesch} 2018}]{KM18}
{Karak}, B.~B., \& {Miesch}, M.~S., 2018, arXiv:1712.10130

\bibitem[{{Passos} {et~al.}(2014)}]{Pas14}
{Passos}, D., {Nandy}, D., {Hazra}, S., \& {Lopes}, I. 2014, \aap, 563, A18

\bibitem[{{Lemerle} \& {Charbonneau}(2017)}]{LC16}
{Lemerle}, A., \& {Charbonneau}, P. 2017, \apj, 834, 133

\bibitem[{{Hazra} {et~al.} 2017}]{HCM17}
{Hazra}, G., {Choudhuri}, A.~R., \& {Miesch}, M.~S. 2017, \apj, 835, 39

\bibitem[{{Stenflo} \& {Kosovichev} 2012}]{SK12}
{Stenflo}, J.~O., \& {Kosovichev}, A.~G. 2012, \apj, 745, 129

\bibitem[{{Mandal} {et~al.} 2017}]{MKB17}
{Mandal}, S., {Karak}, B.~B., \& {Banerjee}, D. 2017, \apj, 851, 70

\bibitem[{McIntosh} {et~al.} 2013]{McInt13}
{McIntosh}, S.~W., {et~al.} 2013, \apj, 765, 146

\bibitem[{{Augustson} {et~al.} 2015}]{ABMT15}
{Augustson}, K., {Brun}, A.~S., {Miesch}, M., \& {Toomre}, J. 2015, \apj, 809,  149

\bibitem[{{Karak} {et~al.} 2015}]{KKB15}
{Karak}, B.~B. and {Kitchatinov}, L.~L. and {Brandenburg}, A. 2015, \apj, 803, 95

\bibitem[{{K{\"a}pyl{\"a}} {et~al.} 2016}]{Kap16}
{K{\"a}pyl{\"a}}, M.~J., {K{\"a}pyl{\"a}}, P.~J., {Olspert}, N., {Brandenburg},  et al.\ 2016, \aap, 589, A56


\end{thebibliography}
\end{document}